\documentstyle[prd,aps]{revtex}

\newcommand{\bea}{\begin{eqnarray}}
\newcommand{\eea}{\end{eqnarray}}

\begin{document}

\draft
\twocolumn[\hsize\textwidth\columnwidth\hsize\csname
@twocolumnfalse\endcsname

\title{Quintessential perturbations during scaling regime}
\author{Jai-chan Hwang${}^{(a,b)}$ and Hyerim Noh${}^{(c,b)}$ \\
        ${}^{(a)}$ Department of Astronomy and Atmospheric Sciences,
                   Kyungpook National University, Taegu, Korea \\
        ${}^{(b)}$ Institute of Astronomy, Madingley Road, Cambridge, UK \\
        ${}^{(c)}$ Korea Astronomy Observatory, Daejon, Korea}
\date{\today}
\maketitle

\begin{abstract}

The scalar field with an exponential potential allows a scaling solution 
where the the density of the field follows the density of the dominating fluid.
Such a scaling regime is often used as an important ingredient in many 
models of quintessence.
We analyse evolution of perturbations while the background follows the scaling.
As the results, the perturbed scalar field also scales with the 
perturbed fluid, and the perturbations accompany the adiabatic 
as well as the isocurvature mode between the fluid and the field.

\end{abstract}


\vskip2pc]

Recent observational advances in high redshift supernovae, and
the small angular-scale CMBR temperature anisotropy had 
spurred renewed interest in the possible acceleration of the
present day universe.
The cosmological constant would be the first available
and historically well studied explanation for such an 
unexpected state of the present universe.
In order to make the universe to start accelerating only in the
latest moment in logarithmic time interval we need high level 
of fine tuning of the amount of the cosmological constant.
In a way to avoid such a fine tuning a new paradigm of simulating
the late acceleration based on a minimally coupled scalar field
has been proposed, with such a field termed a quintessence.
Although the prime motivation of reducing the fine tuning problem
has not been quite successful, we notice a variety of
roles the field with different potential could achieve 
diverse evolutions of the world model
\cite{Lucchin-Matarrese-1985,%
Halliwell-1987,%
Ratra-Peebles-1988,%
Wetterich-1988,%
Wetterich-1995,%
Caldwell-etal-1998,%
Kolda-Lyth-1998,%
Peebles-Vilenkin-1998,%
Steinhardt-etal-1998,%
Barreiro-etal-1999,%
Efstathiou-1999,%
Peebles-1999,%
Sahni-Wang-1999,%
Bean-Magueijo-2000,%
Dimopoulos-2000,%
Dodelson-etal-2000,%
Hebecker-Witterich-2000,%
Weller-Albrecht-2000,%
Zimdahl-Pavon-2001}.

It has been known that an exponential potential supports a scaling 
attractor regime where the density of the field follows the
density of the dominating fluid in the background 
\cite{Lucchin-Matarrese-1985,%
Halliwell-1987,%
Ratra-Peebles-1988,%
Wetterich-1988}.
Although the scaling by itself cannot explain the acceleration
in the present epoch, since the scaling works as an attractor
we find many implementation of quintessence idea
having such an attractor in the early evolution stage.
The evolution of perturbation during the scaling regime as well
as the following quintessential development has been
actively studied in recent literature 
\cite{Ratra-Peebles-1988,%
Wetterich-1995,%
Coble-etal-1996,%
Ferreira-Joyce-1998,%
Perrotta-Baccigalupi-1998,%
Viana-Liddle-1998,%
Wang-Steinhardt-1998,%
Perrotta-etal-1999,%
Amendola-2000,%
Baccigalupi-etal-2000,%
Brax-etal-2000,%
Doran-etal-2000,%
Skordis-Albrecht-2000,%
Abramo-Finelli-2001,%
Bean-2001,%
Kawasaki-etal-2001,%
Ott-2001}.

In this paper we study the evolution of perturbations 
in a fluid-field system with an exponential field potential.
We will show that the conventional growing and decaying solutions
of the fluid without the field {\it remain valid} with two additional
decaying solutions due to the coupling, 
eqs. (\ref{solution-1},\ref{solution-m1},\ref{solution-w1}).
The {\it perturbed scalar field also scales} the perturbed density
field of the fluid,
eqs. (\ref{solution-2},\ref{solution-m2},\ref{solution-w2}).
As a result we show that while the background evolution follows
the scaling solution the perturbations {\it accompany} both the adiabatic 
and the isocurvature modes,
eqs. (\ref{solution-3},\ref{solution-m3},\ref{solution-w3}).
In order to make the paper self contained, and for future convenient usage,
the notations and basic set of equations are summarized in the Appendix.

\vskip .5cm
{\it Perturbed equations:}
The pressureless matter ($c$) or the radiation ($r$) dominated eras in three
component system ($r$, $c$ and a scalar field $\phi$) 
can be handled effectively based on a 
two component system composed of a scalar field and an ideal fluid ($f$) with
$w \equiv p_{f}/\mu_{f} = {\rm constant}$, see below.
As the gauge condition we take $v_{f} = 0$ which can be regarded
as the comoving gauge based on the dominating fluid component.
Equations (\ref{G6},\ref{G7},\ref{G-MSF},\ref{G5}) lead to
\bea
   & & \ddot \delta_{f} + \left( 2 - 3 w \right) H \dot \delta_{f}
       + \Big[ w {k^2 \over a^2}
       - 6 w \left( \dot H + H^2 \right)
   \nonumber \\
   & & \qquad
       - 4 \pi G \mu_{f} (1 + w) (1 + 3 w) \Big] \delta_{f}
   \nonumber \\
   & & \qquad
       = 8 \pi G \left[ 2 w \dot \phi^2 \delta_{f}
       + (1 + w) \left( 2 \dot \phi \delta \dot \phi
       - V_{,\phi} \delta \phi \right) \right],
   \label{eq1} \\
   & & \delta \ddot \phi + 3 H \delta \dot \phi
       + \Big( {k^2 \over a^2} + V_{,\phi\phi} \Big) \delta \phi
   \nonumber \\
   & & \qquad
       = {1 \over 1 + w} \left[ (1- w) \dot \phi \dot \delta_{f}
       - 2 w \left( \ddot \phi + 3 H \dot \phi \right)
       \delta_{f} \right].
   \label{eq2}
\eea
The LHS of eq. (\ref{eq1}) is the familiar density perturbation equation
in the comoving gauge derived in \cite{Nariai-1969,Bardeen-1980},
and the LHS of of eq. (\ref{eq2}) is the familiar perturbed scalar field
equation without the metric perturbation; compare with eq. (\ref{G-MSF}).
Thus, our gauge choice allows simple equations for the
uncoupled parts of the system.
However, eq. (\ref{eq2}) is not valid in a situation without the fluid;
in such a case we should go back to the original set of equation 
in a gauge-ready form in the Appendix.

We emphasize that, since $v_f = 0$ gauge condition fixes the temporal gauge
mode completely, the remaining variables in this gauge are all 
gauge-invariant.
{}For example, in our gauge condition we have
$\delta_f = \delta_{f v_f}$ and $\delta \phi = \delta \phi_{v_f}$ where
\bea
   & & \delta_{f v_f} \equiv \delta_f
       + 3 ( 1 + w ) {aH \over k} v_f, \quad
       \delta \phi_{v_f} \equiv \delta \phi - \dot \phi {a \over k} v_f.
\eea
In our notation, subindices $f, c, r, \phi$ indicate the fluid or field
component, whereas the others with perturbed order variables indicate
the gauge-invariant notation.
{}For our convention of the gauge-invariant combinations
see \S 3 in \cite{Hwang-1991-PRW}.

Under our gauge condition eq. (\ref{G7}) gives
$\alpha = - {w \over 1 + w} \delta_{f}$, thus {\it for} a pressureless
fluid $v_{f} = 0$ gauge implies $\alpha = 0$ which is the synchronous 
gauge condition; the opposite is not necessarily true.
In a more realistic situation we should consider a system with 
a radiation ($r$) a pressureless matter ($c$) and a scalar field ($\phi$).
Many of the previous works prefered to use the synchronous gauge.
Equation (\ref{G7}) gives $(a v_{c})^\cdot = k \alpha$, 
thus, $v_{c} \equiv 0$ implies the synchronous gauge condition; 
by further ignoring $v_{c}$, which otherwise will give the gauge mode, 
the synchronous gauge becomes effectively the same as the $v_{c} = 0$ gauge.
{}From eqs. (\ref{G6},\ref{G7},\ref{G-MSF},\ref{G5}) we can derive 
the set of equations for the three component system under the
$v_{c} = 0$ gauge.
In the radiation-less limit such equations will
produce correctly $w = 0$ limit of eqs. (\ref{eq1},\ref{eq2});
this is the case for $\delta_{r} \sim \delta_{c}$.
Without the pressureless fluid, however, such three-component equations
will {\it not} produce $w = {1 \over 3}$ limit of eqs. (\ref{eq1},\ref{eq2});
the difference is due to the different gauge conditions used.
Since the $v_{c} = 0$ gauge is based on the pressureless matter, 
without or negligible amount of the pressureless matter 
(as in the radiation dominated era, RDE) 
we could encounter situations where the gauge condition is 
{\it not suitable} for handling the problem; the situation is not incorrect
as long as we have a tiny fraction of the 
pressureless matter in the RDE, but later we will notice a situation
(in the literature) where it (the synchronous gauge) causes more 
troublesome analytic handling compared with analyses based on our 
present gauge condition.

\vskip .5cm
{\it Scaling background:}
It is known that with a potential $V = V_0 e^{- \lambda \phi}$ 
the scalar field approaches an attractor solution of the dominant component
\cite{Lucchin-Matarrese-1985,Halliwell-1987,Ratra-Peebles-1988,Wetterich-1988}.
We set $8 \pi G \equiv 1$.
In a flat background without $\Lambda$, if the energy density of dominant 
component ($f$) behaves as $\mu_{f} \propto a^{-3(1+w)}$,
eqs. (\ref{BG1}-\ref{BG3}) allow
\cite{Liddle-Scherrer-1998,Ferreira-Joyce-1998,Skordis-Albrecht-2000}:
\bea
   & & \mu_{\phi} = {2 \over 1 - w} V = {1 \over 1 + w} \dot \phi^2, \quad
       \Omega_{\phi} = {3(1+w) \over \lambda^2}, 
   \nonumber \\
   & & \phi^\prime = {3(1+w) \over \lambda} {a^\prime \over a}
       = {6(1+w) \over 1+3w} {1 \over \lambda \eta},
   \label{scaling}
\eea
and $\Omega_{f} = 1 - \Omega_{\phi}$;
a prime indicates a time derivative based on $\eta$.
With this we can show $p/\mu = w = p_{\phi}/\mu_{\phi}$ as well.
In the following we analyse the perturbations for
the cases of a radiation dominated era, a matter dominated era,
and an era with general $w$.

\vskip .5cm
{\it Case with $w = {1 \over 3}$:}
When the dominant component is the radiation, eqs. (\ref{eq1},\ref{eq2}) give
\bea
   & & \delta_{r}^{\prime\prime} + \Big[ {k^2 \over 3}
       - \Big( 1 - {8 \over 3 \lambda^2} \Big) {2 \over \eta^2}
       \Big] \delta_{r}
       = {16 \over 3 \lambda} \Big( {2 \over \eta} \delta \phi^\prime
       + {1 \over \eta^2} \delta \phi \Big),
   \label{scaling-r-delta-eq} \\
   & & \delta \phi^{\prime\prime} + {2 \over \eta} \delta \phi^\prime
       + \Big( k^2 + {4 \over \eta^2} \Big) \delta \phi
       = {2 \over \lambda} \Big( {1 \over \eta} \delta_{r}^\prime
       - {1 \over \eta^2} \delta_{r} \Big).
   \label{scaling-r-phi-eq}
\eea
In the large-scale limit, ignoring $k^2$ terms, we have solutions
$\delta_{r} \propto \delta \phi \propto \eta^q$ where
\bea
   & & q = 2, \quad -1, \quad
       {1 \over 2} \left( -1 \pm \sqrt{1 - 16 \Omega_{r}} \right).
   \label{solution-1}
\eea
These provide the {\it general solutions} for $\delta_{r}$ and $\delta \phi$.
Notice that the first two solutions are the well known
growing and decaying modes of $\delta_{r}$ in the absence of the field
\cite{Nariai-1969}.
{}For the growing solution $\delta_{r} \propto \eta^2$ we have
\bea
   & & \delta_{r} = {5 \over 2} \delta_{\phi} = 5 \lambda \delta \phi.
   \label{solution-2}
\eea
We can show $\delta \equiv \delta \mu/\mu
= \Omega_{r} \delta_{r} + \Omega_{\phi} \delta_{\phi}
= ( 1 - {3 \over 5} \Omega_{\phi} ) \delta_{r}$, and
$\delta_v = \delta + 3H \dot \phi \delta \phi/\mu$
where $\delta_v \equiv \delta + 3 (1 + w) (aH / k) v$.
Thus, for the growing solution we have
\bea
   & & \delta_v = \Big( 1 - {2 \over 5} \Omega_{\phi} \Big) \delta_{r},
       \quad
       S_{r\phi} = {9 \over 20} \delta_{r},
   \label{solution-3}
\eea
where \cite{KS-1984}
\bea
   & & S_{ij} \equiv {\delta \mu_{i} \over \mu_{i} + p_{i}}
       - {\delta \mu_{j} \over \mu_{j} + p_{j}}.
\eea
The adiabatic and the isocurvature perturbations are often characterized by
$\delta_v$ and $S_{ij}$, respectively.
{}From eqs. (\ref{G2},\ref{G3}) we have
\bea
   & & {k^2 - 3K \over a^2} \varphi_\chi = {1 \over 2} \mu \delta_v,
   \label{Poisson-eq}
\eea
where $\varphi_\chi \equiv \varphi - H \chi$; in the notation of
\cite{Bardeen-1980,MFB-1992} we have $\varphi_\chi = \Phi_H = - \Psi$. 
Thus, vanishing $\delta_v$ implies vanishing $\varphi_\chi$, 
thus isocurvature in the multi-component situation. 
Therefore, eq. (\ref{solution-3}) shows that the growing mode of 
perturbation during the scaling regime accompanies the adiabatic ($\delta_v$)
as well as the isocurvature ($S_{r \phi}$) modes;
this is {\it in contrast} with the assumption made in \cite{Viana-Liddle-1998},
and the result in \cite{Abramo-Finelli-2001}\footnote{
         The work in \cite{Abramo-Finelli-2001} was made in the 
         zero-shear gauge which fixes the scalar-type shear of the
         normal hypersurface $\chi$ equal to zero 
         \cite{Bardeen-1980,Hwang-1991-PRW}.
         $\delta_{f}$ in the zero-shear gauge is the same as
         $\delta_{f\chi} \equiv \delta_{f} + 3 (1 + w) H \chi$.
         In our gauge condition we can
         show that $H \chi$ is of the order $\delta_{f}/(k\eta)^2$. 
         Thus, in the zero-shear gauge, the shear (thus metric) part
         dominates the density fluctuation in the large-scale limit. 
         Although the authors of \cite{Abramo-Finelli-2001}
         concluded that they found no isocurvature 
         mode during scaling, what they actually have shown was that 
         to the order of perturbed potential $\varphi_\chi$ 
         the isocurvature mode vanishes.
         What we have shown is that we have an accompanied isocurvature mode
         which is of the order of denisty perturbation $\delta_v
         = {2 \over 3} ({k \over aH})^2 \varphi_\chi$.
         In order to derive our result in the zero-shear gauge
         one has go to higher order terms in the large-scale expansion.
         }.

Results in \cite{Ferreira-Joyce-1998} (see also 
\cite{Skordis-Albrecht-2000,Amendola-2000}) also {\it differ} from ours 
in eqs. (\ref{solution-1}-\ref{solution-3})\footnote{
         In this context, compare with eq. (44) and the solutions below this
         equation in \cite{Ferreira-Joyce-1998},
         eqs. (23-29) in \cite{Skordis-Albrecht-2000},
         and eqs. (49,50) in \cite{Amendola-2000}.
         The work in \cite{Brax-etal-2000} ignored the connections between 
         the quintessence field and the fluid: this applies to eqs. (30,43) 
         and the subsequent results in \cite{Brax-etal-2000}.
         }.
We do not expect the results in eqs. (\ref{solution-1},\ref{solution-2}) 
should be the same because the gauges used are different.
The growing mode {\it happens} to coincide with ours in eq. (\ref{solution-1}), 
but notice that the relation between the growing solutions in
eq. (\ref{solution-2}) differs; compare with results below eq. (44)
in \cite{Ferreira-Joyce-1998} and eq. (29) in \cite{Skordis-Albrecht-2000}.
However, since $\delta_v$ and $S_{r\phi}$ are gauge-invariant
the relation between them 
\bea
   & & \delta_v = {20 \over 9} \Big( 1 - {2 \over 5} \Omega_{\phi} \Big)
       S_{r\phi},
   \label{relation}
\eea
should be the same.
We can derive the same solution using the fluid formulation of 
the perturbed field system; i.e., using eqs. (\ref{G6},\ref{G7},\ref{e_phi})
for the field, see below eq. (\ref{G-MSF}). 
However, if we simply follow the analyses in 
\cite{Ferreira-Joyce-1998,Skordis-Albrecht-2000,Amendola-2000,Brax-etal-2000}
it is not easy to get our results.
The works in \cite{Ferreira-Joyce-1998,Skordis-Albrecht-2000,Amendola-2000}
are based on the synchronous gauge which is effectively
the comoving gauge based on the pressureless matter.
Since the pressureless matter is subdominating in RDE we anticipate 
that the gauge condition is {\it not} the best one to suit the situation.
Equation (\ref{solution-3}) is not presented in 
\cite{Ferreira-Joyce-1998,Skordis-Albrecht-2000,Amendola-2000}, 
and one {\it cannot} simply derive eqs. (\ref{solution-3},\ref{relation}) 
based on their solutions; to get the correct answer in their synchronous gauge 
condition one probably needs to consider higher order solutions in the 
large-scale expansion or has to consider the full three-component
system with the pressureless component.
We do not need such extra efforts in our gauge condition.
By handling the three-component system in the synchronous 
(more precisely, $v_{c} = 0$) gauge condition, recently, 
we have derived the above solutions successfully, \cite{Carsten-Weller-2001}.

In the small-scale limit eqs. (\ref{eq1},\ref{eq2}) give two solutions
$\delta_{r} \propto e^{\pm i k \eta /\sqrt{3}}$ which are the same
solutions valid for pure radiation \cite{Nariai-1969}, 
and the remaining two solutions decay faster.

\vskip .5cm
{\it Case with $w = 0$:}
In the large-scale limit eqs. (\ref{eq1},\ref{eq2}) have solutions
$\delta_{c} \propto \delta \phi \propto \eta^q$ where
\bea
   & & q = 2, \quad -3, \quad
       {3 \over 2} \left( -1 \pm \sqrt{ 1 - 8 \Omega_{c}} \right).
   \label{solution-m1}
\eea
The first two solutions are the well known
growing and decaying modes of $\delta_{c}$ in the absence of the field
\cite{Lifshitz-1946,Nariai-1969}.
{}For the dominant solution $\delta_{c} \propto \eta^2$ we have
\bea
   & & \delta_{c} = - 14 \delta_{\phi} = {7 \over 3} \lambda \delta \phi.
   \label{solution-m2}
\eea
We can show $\delta = ( 1 - {15 \over 14} \Omega_{\phi} ) \delta_{c}$,
thus,
\bea
   & & \delta_v = \Big( 1 - {9 \over 14} \Omega_{\phi} \Big) \delta_{c}, 
       \quad
       S_{c\phi} = {15 \over 14} \delta_{c}.
   \label{solution-m3}
\eea
These are the adiabatic and isocurvature fractions accompanied by the
growing mode.
{}For $w = 0$ our gauge condition coincides with the synchronous gauge.
However, eq. (\ref{solution-m3}) was not presented previously,
and somehow results in eqs. (\ref{solution-m1},\ref{solution-m2})
differ from the ones in \cite{Ferreira-Joyce-1998};
compare with results below eq. (45) in \cite{Ferreira-Joyce-1998}
which are in errors.

In the small-scale limit we have two solutions
$\delta_{c} \propto \eta^q$ where 
$q = {1 \over 2} ( - 1 \pm \sqrt{1 + 24 \Omega_{c}} )$,
which lead to $q = 2, -3$ for vanishing $\Omega_{\phi}$,
\cite{Ferreira-Joyce-1998}.
This is the well known perturbation growth rate in the presence of 
unclustered component of matter derived in 
\cite{Bond-etal-1980,Hoffman-Bludman-1984}.
We can show that additional two solutions decay faster.

\vskip .5cm
{\it Case with general $w$:}
Now, we consider the general situation with constant $w$.
In the large-scale limit we have solutions
$\delta_{f} \propto \delta \phi \propto \eta^q$ with
\bea
   & & q = 2, \quad 
       - {3 (1 - w) \over 1 + 3 w}, 
   \nonumber \\
   & & \qquad
       {3 (1 - w) \over 2(1 + 3 w)} \Big( - 1 
       \pm \sqrt{ 1 - 8 {1 + w \over 1 - w} \Omega_{f} } \Big).
   \label{solution-w1}
\eea
These give the general solutions for $\delta_{f}$ and $\delta \phi$.
The first two solutions are the well known
growing and decaying modes of $\delta_{f}$ in the absence of the field:
$\delta_{f} \propto a^{1+3w}$ and $a^{-{3 \over 2} (1 - w)}$
\cite{Nariai-1969,Bardeen-1980}.
{}For the dominant solution $\delta_{f} \propto \eta^2$ we have
\bea
   & & \delta_{f} = - {2(7 + 9 w) \over 1 - 24 w - 9w^2} \delta_{\phi} 
        = {7 + 9 w \over 3 ( 1 - w)} \lambda \delta \phi.
   \label{solution-w2}
\eea
{}Following the similar calculations as before we have
\bea
   & & \delta_v = \left[ 1 - {9(1-w^2) \over 2(7+9w)} \Omega_{\phi}
       \right] \delta_{f}, 
   \nonumber \\
   & & S_{f\phi} = { 3 (1 - w) (5 + 3 w) \over 2 (1 + w)(7 + 9w) }
       \delta_{f},
   \label{solution-w3}
\eea
which show the adiabatic and isocurvature modes respectively accompanied by
the growing mode of perturbation during the scaling regime.

It is convenient to properly normalize the coefficient with the
well known conserved quantity in the large-scale (super-sound-horizon scale)
limit: $\varphi_v \equiv \varphi - (aH/k) v$.
{}For the growing solution we have
$\varphi_v = {5 + 3 w \over 3 + 3 w} \varphi_\chi$.
Thus, using eq. (\ref{Poisson-eq}) we have
\bea
   & & \varphi_v = { 2 (5 + 3 w) \over (1 + w)(1 + 3w)^2 }
       \left[ 1 - {9(1-w^2) \over 2(7+9w)} \Omega_{\phi} \right] 
       {1 \over (k \eta)^2} \delta_{f}.
   \nonumber \\
\eea
In the large-scale limit we have $\varphi_v = C({\bf x})$ which
will fix the normalizations of $\delta_{f}$ and others in terms of $C$.

\vskip .5cm
{\it Summary:}
We have considered a system of a field and a fluid with constant $w$.
The system includes limits of a three component system ($r$, $c$, $\phi$)
when either $r$ or $c$ is dominating over the other.
We find that while the field scales with the fluid in the background
the perturbed field also scales with the perturbed fluid,
eqs. (\ref{solution-2},\ref{solution-m2},\ref{solution-w2}).
The ordinary growing and decaying solutions
of the fluid without the field remain valid for the coupled system,
eqs. (\ref{solution-1},\ref{solution-m1},\ref{solution-w1}).
The main results of our work are
eqs. (\ref{solution-3},\ref{solution-m3},\ref{solution-w3})
which show that during the scaling the perturbations accompany 
both the adiabatic and the isocurvature modes.
Considering the fact that a perturbed field accompanies 
nonvanishing entropic perturbation \cite{BST-1983},
$e_\phi \equiv \delta p_\phi - c_\phi^2 \delta \mu_\phi$,
it may not be a surprise to see the accompanied isocurvature mode
in the multi-component system with the field.

\subsection*{Acknowledgments}

We thank Martin Bucher and George Efstathiou for helpful discussions.
We also wish to thank Carsten van de Bruck, Neil Turok and Jochen Weller
for making useful suggestions and remarks.
HN was supported by grant No. 2000-0-113-001-3 from the
Basic Research Program of the Korea Science and Engineering Foundation.
JH was supported by the Korea Research Foundation Grants 
(KRF-2000-013-DA004 and 2000-015-DP0080).

\section*{Equations of fluid-field system}
                                   \label{sec:eqs}

\setcounter{equation}{0}
\def\theequation{A\arabic{equation}}

We consider a system of a multiple ideal fluids and a field
without direct mutual interactions among components; see
\cite{Hwang-Noh-2001-Fluids} for more general situations.
Our convension of the metric and the energy-momentum tensor is:
\bea
   & & d s^2 = - a^2 \left( 1 + 2 \alpha \right) d \eta^2
       - 2 a^2 \beta_{,\alpha} d \eta d x^\alpha
   \nonumber \\
   & & \qquad
       + a^2 \left[ g^{(3)}_{\alpha\beta} \left( 1 + 2 \varphi \right)
       + 2 \gamma_{,\alpha|\beta} \right] d x^\alpha d x^\beta,
   \label{metric-general} \\
   & & T^0_0 = - \left( \bar \mu + \delta \mu \right), \quad
       T^0_\alpha = - {1 \over k} \left( \mu + p \right) v_{,\alpha},
   \nonumber \\
   & & T^\alpha_\beta = \left( \bar p + \delta p \right) \delta^\alpha_\beta
       + \Big( {1 \over k^2} \nabla^\alpha \nabla_\beta
       + {1 \over 3} \delta^\alpha_\beta \Big) \pi^{(s)}.
   \label{Tab}
\eea
$\beta$ and $\gamma$ always appear together as
$\chi \equiv a ( \beta + a \dot \gamma )$ which is spatially gauge-invariant.
An overdot denotes a time
derivative based on $t$ with $dt \equiv a d \eta$.
$\bar \mu = \sum_{l} \bar \mu_{l}$, $\delta \mu = \sum_{l} \delta \mu_{l}$,
and similarly for $\bar p$, $\delta p$, $(\mu + p) v$ and $\pi^{(s)}$.
The sum includes the field contribution which
can be expressed in terms of fluid variables:
\bea
   & & \mu_{\phi} \equiv {1 \over 2} \dot \phi^2 + V, \quad
       \delta \mu_{\phi} \equiv \dot \phi \delta \dot \phi
       - \dot \phi^2 \alpha + V_{,\phi} \delta \phi, 
   \nonumber \\
   & & p_{\phi} \equiv {1 \over 2} \dot \phi^2 - V, \quad
       \delta p_{\phi} \equiv \dot \phi \delta \dot \phi
       - \dot \phi^2 \alpha - V_{,\phi} \delta \phi,
   \nonumber \\
   & & v_{\phi} \equiv {k \over a} { \delta \phi \over \dot \phi },
       \quad
       \pi_{\phi}^{(s)} \equiv 0.
   \label{Fluid-phi}
\eea

The background evolution is governed by
\bea
   & & H^2 = {8 \pi G \over 3} \mu + {\Lambda \over 3} - {K \over a^2}, \quad
   \label{BG1} \\
   & & \dot \mu_{i} + 3 H \left( \mu_{i} + p_{i} \right) = 0,
   \label{BG2} \\
   & & \ddot \phi + 3 H \dot \phi + V_{,\phi} = 0.
   \label{BG3}
\eea
The energy-momentum conservation of individual fluid and
the equation of motion of a scalar field give \cite{Hwang-Noh-2001-Fluids}
\bea
   & & \delta \dot \mu_{i}
       + 3 H \left( \delta \mu_{i} + \delta p_{i} \right)
       = \left( \mu_{i} + p_{i} \right) \Big( \kappa
       - 3 H \alpha - {k \over a} v_{i} \Big),
   \nonumber \\
   \label{G6} \\
   & & {\left[ a^4 ( \mu_{i} + p_{i} ) v_{i} \right]^\cdot
       \over a^4 ( \mu_{i} + p_{i} )}
       = {k \over a} \Big( \alpha
       + {\delta p_{i} \over \mu_{i} + p_{i}} \Big), 
   \label{G7} \\
   & & \delta \ddot \phi + 3 H \delta \dot \phi
       + {k^2 \over a^2} \delta \phi
       + V_{,\phi\phi} \delta \phi
   \nonumber \\
   & & \qquad
       = \dot \phi \left( \kappa + \dot \alpha \right)
       + \left( 2 \ddot \phi + 3 H \dot \phi \right) \alpha,
   \label{G-MSF}
\eea
where $\delta_{i} \equiv \delta \mu_{i} / \mu_{i}$,
$w_{i} \equiv p_{i} / \mu_{i}$, and
$c_{i}^2 \equiv \dot p_{i} / \dot \mu_{i}$.
Equations (\ref{G6},\ref{G7}) remain valid for the scalar field
with $i = \phi$. 
{}From eq. (\ref{Fluid-phi}) we can show
\bea
   & & \delta p_\phi = \delta \mu_\phi
       + 3 {aH \over k} ( 1 - c_\phi^2 ) ( \mu_\phi + p_\phi ) v_\phi.
   \label{e_phi}
\eea
The metric parts are described by Einstein's equations
\cite{Bardeen-1988,Hwang-1991-PRW}:
\bea
   & & \kappa \equiv 3 \left( - \dot \varphi + H \alpha \right)
          + {k^2 \over a^2} \chi,
   \label{G1} \\
   & & - {k^2 - 3K \over a^2} \varphi + H \kappa
       = - 4 \pi G \delta \mu,
   \label{G2} \\
   & & \kappa - {k^2 - 3 K \over a^2} \chi
       = 12 \pi G {a \over k} ( \mu + p ) v,
   \label{G3} \\
   & & \dot \chi + H \chi - \alpha - \varphi
       = 8 \pi G {a^2 \over k^2} \pi^{(s)},
   \label{G4} \\
   & & \dot \kappa + 2 H \kappa
       + \Big( 3 \dot H - {k^2 \over a^2} \Big) \alpha
       = 4 \pi G \Big( \delta \mu + 3 \delta p \Big).
   \label{G5}
\eea
These equations are in a gauge-ready form. 
That is, we have not chosen the temporal gauge condition so that we 
could {\it use the freedom as an advantage} in handling problems in 
diverse situations; the variables are all spatially gauge-invariant 
\cite{Bardeen-1988,Hwang-1991-PRW}.
More general equations with multiple fluids and fields with general
interactions among them can be found in \cite{Hwang-Noh-2001-Fluids}

\baselineskip 0pt

\end{document}